\documentclass[prd,superscriptaddress,amsfonts,amssymb,amsmath,showpacs,twocolumn]{revtex4}
    \usepackage{epsfig}
    \usepackage{graphics}  
    \usepackage{color}
    \date{\today}

\begin{document}

    \title{Classical and quantum aspects of the radiation emitted 
by a uniformly accelerated charge: Larmor--Unruh reconciliation 
and zero-frequency Rindler modes} 



    \author{Andr\'e G.\ S.\ Landulfo}\email{andre.landulfo@ufabc.edu.br}
    \affiliation{Centro de Ci\^encias Naturais e Humanas,
    Universidade Federal do ABC,
    Avenida dos Estados, 5001, 09210-580,
    Santo Andr\'e, S\~ao Paulo, Brazil}
 
    \author{Stephen A. Fulling}\email{fulling@math.tamu.edu}    
    \affiliation{Department of Physics \& Astronomy, Texas A\&M University, College Station, Texas, 77843-4242, USA}  
    \affiliation{Department of Mathematics, Texas A\&M University, College Station, Texas, 77843-3368, USA}
    
    \author{George E.\ A.\ Matsas}\email{matsas@ift.unesp.br}
    \affiliation{Instituto de F\'\i sica Te\'orica, Universidade
    Estadual Paulista, Rua Dr.\ Bento Teobaldo Ferraz, 271, 01140-070,
    S\~ao Paulo, S\~ao Paulo, Brazil}


    \begin{abstract}
    The interplay between acceleration and radiation harbors remarkable and surprising consequences. 
    One of the most striking is that the Larmor radiation emitted by a charge can be seen as a consequence of
    the Unruh thermal bath.  Indeed, this connection between the Unruh effect and classical bremsstrahlung was 
    used recently to propose an experiment to  confirm  (as directly as possible) the existence of the Unruh thermal bath.  
    This situation may sound puzzling in two ways:  first,  because the Unruh effect is a strictly
    quantum effect while Larmor radiation is a classical one, and second, because of  the crucial role played by zero-frequency 
    Rindler photons in this context. In this paper we settle these two issues by showing how the quantum evolution leads naturally 
    to all relevant aspects of Larmor radiation and, especially,  how the corresponding  classical radiation is entirely built from 
    such zero-Rindler-energy modes.    
 \end{abstract}

    \maketitle

\section{Introduction}
\label{sec:I}

The connection between acceleration and radiation is a subject that repeatedly  puzzles
physicists. It is well known (e.g.,~\cite{ZW})  that an accelerated charge radiates 
electromagnetic waves (bremsstrahlung) in the inertial frame with an emitted power 
given by the famous Larmor formula~\cite{Larmor}. However, ever since the works of 
Rohrlich~\cite{R1, R2} and Boulware~\cite{B} it has been known that radiation is not a local, 
covariant concept. They found that although a uniformly accelerated electric charge radiates with 
respect to distant inertial observers, uniformly co-accelerated observers see no radiation coming 
from the charge.   
   
More recently,  this issue was investigated within the framework of quantum field theory (QFT) in curved 
spacetimes~\cite{HMS92} (see also Refs.~\cite{HMS92a, HM, RW,  PV}).  This work 
found what appears to be a striking connection between  bremsstrahlung and a remarkable effect 
of QFT discovered by Unruh in 1976~\cite{U}. The Unruh effect states that 
uniformly accelerated observers with proper acceleration $a$ detect a thermal bath of particles with 
temperature
\begin{equation}
T_U=\hbar a/(2\pi c k_B)
\label{TU}
\end{equation} 
when the quantum field is in the Minkowski vacuum. By analyzing the bremsstrahlung effect using 
QFT in both inertial and co-accelerated frames, it was shown~\cite{HMS92}  
that co-accelerating  observers with the uniformly accelerated charge interpret the usual (inertial) 
emission as the combined rate of emission and absorption of {\em zero-energy} Rindler photons 
(energy defined with respect to Rindler time) into and from the Unruh thermal bath, respectively. 
The connection between the Unruh effect and the bremsstrahlung was strengthened recently in  
Ref.~\cite{CLMV17},  where  it was shown that the existence of the Unruh effect is reflected 
in the {\em classical} Larmor radiation emitted by an accelerated charge. This connection was used 
to propose an experiment reachable under present technology whose result may be directly 
interpreted in terms of the Unruh thermal bath (see also Ref.~\cite{CLMV18} for more details on 
the experiment). 

This body of work has left two important issues not fully resolved. The first one is the crucial role 
played in the QFT calculations  by {\em zero-energy} Rindler photons, which (in the limit of literally 
zero Rindler frequency) are  pressed completely into the boundary (horizon) of the Rindler wedge 
(see Fig.~\ref{fig1}). The very idea of a nontrivial zero-frequency mode is unfamiliar and gives the
entire argument a mysterious air, which has hindered its acceptance.  Indeed, trenchant
questions and criticisms by D. N. Page and W. G. Unruh (private communications)
persuaded us of the need to perform the present deeper investigation and exposition.
The second issue is how the Unruh temperature~(\ref{TU}),  a manifestly quantum effect,
 becomes  codified in the (classical) Larmor radiation emitted by the charge. In the present paper, 
 we address these two issues by analyzing both classical and quantum aspects of the problem in 
 a way that connects the radiation seen by inertial observers with the physics of uniformly 
 accelerated ones. For the sake of simplicity, we will focus on the radiation emitted by a 
 scalar source, since the above issues are already present in such a case. We first show that 
 zero-energy Rindler modes are the ones responsible for building the radiation seen by inertial 
 observers in the classical scenario. Next, we tackle the problem from the perspective of QFT 
 and show that the quantum evolution takes the radiation field state from the vacuum of the 
 inertial observes in the asymptotic past to a coherent state (with only zero-Rindler-energy 
 excitations contributing to it)  for inertial observes in the asymptotic future. The field 
 expectation value in such a state is given by the classical retarded solution,
 and the (normal ordered) stress-energy tensor expectation value coincides with  its classical 
 counterpart. As a side effect, this paper should also boost the interest of experimentalists 
 to carry on the proposal of observing the Unruh effect raised in Ref.~\cite{CLMV17}.

The paper is organized as follows. In Sec.~\ref{sec:II}, we discuss the classical aspects of 
the radiation emitted by the source and the role played by zero-energy Rindler modes in 
that context. In Sec.~\ref{sec:III} we delve into the quantum evolution of the system and show 
how it relates to the classical analysis. Our closing remarks appear in Sec.~\ref{sec:IV}. 

We adopt metric signature $(-,+,+,+)$ and units where $G =\hbar=c=k_B=1$, unless stated otherwise.

\section{Radiation emitted by an accelerated charge: Classical aspects}
\label{sec:II}

Let us begin by considering a uniformly accelerated scalar source $j$  interacting with a classical scalar 
field $\phi$ for a finite proper time $T_{\rm tot}\equiv 2T$  in Minkowski spacetime $(\mathbb{R}^4, \eta_{ab})$, 
$\eta_{ab}$ being the flat Minkowski metric tensor. The worldline of a uniformly accelerating pointlike source  
is, without loss of generality,
\begin{equation}
t=a^{-1}\sinh a\tau, \; z=a^{-1}\cosh a\tau, \: x=0, \; y=0,
\label{worldline}
\end{equation}
where $\tau$ and  $a$ are the source's proper time and acceleration, respectively, and here $(t,x,y,z) $ are 
the usual Cartesian coordinates covering Minkowski spacetime. The scalar source is then given by~\cite{footnote}
\begin{equation}
j=\left\{\begin{array}{cc} q\delta(\xi)\delta^2({\bf x}_\bot) & -T< \tau <T  \\ 0 & |\tau|> T\end{array}\right. , \quad q={\rm const}
\label{source}
\end{equation}
where we recall that in Rindler coordinates $(\tau, \xi, {\bf x}_\bot)$, $\tau,\xi \in \mathbb{R}$ and ${\bf x}_\bot=(x,y)\in \mathbb{R}^2$,  
covering the right Rindler wedge (region $z>|t|$), the Minkowski line element takes the form
\begin{equation}
ds^2=e^{2a\xi}\left( -d\tau^2 + d\xi^2 \right) + dx^2 + dy^2
\label{rindlermetric}
\end{equation}
 and the worldline~(\ref{worldline}) is cast as $\xi=x=y=0$. 

\begin{figure}
\begin{center}
\includegraphics[scale=0.4]{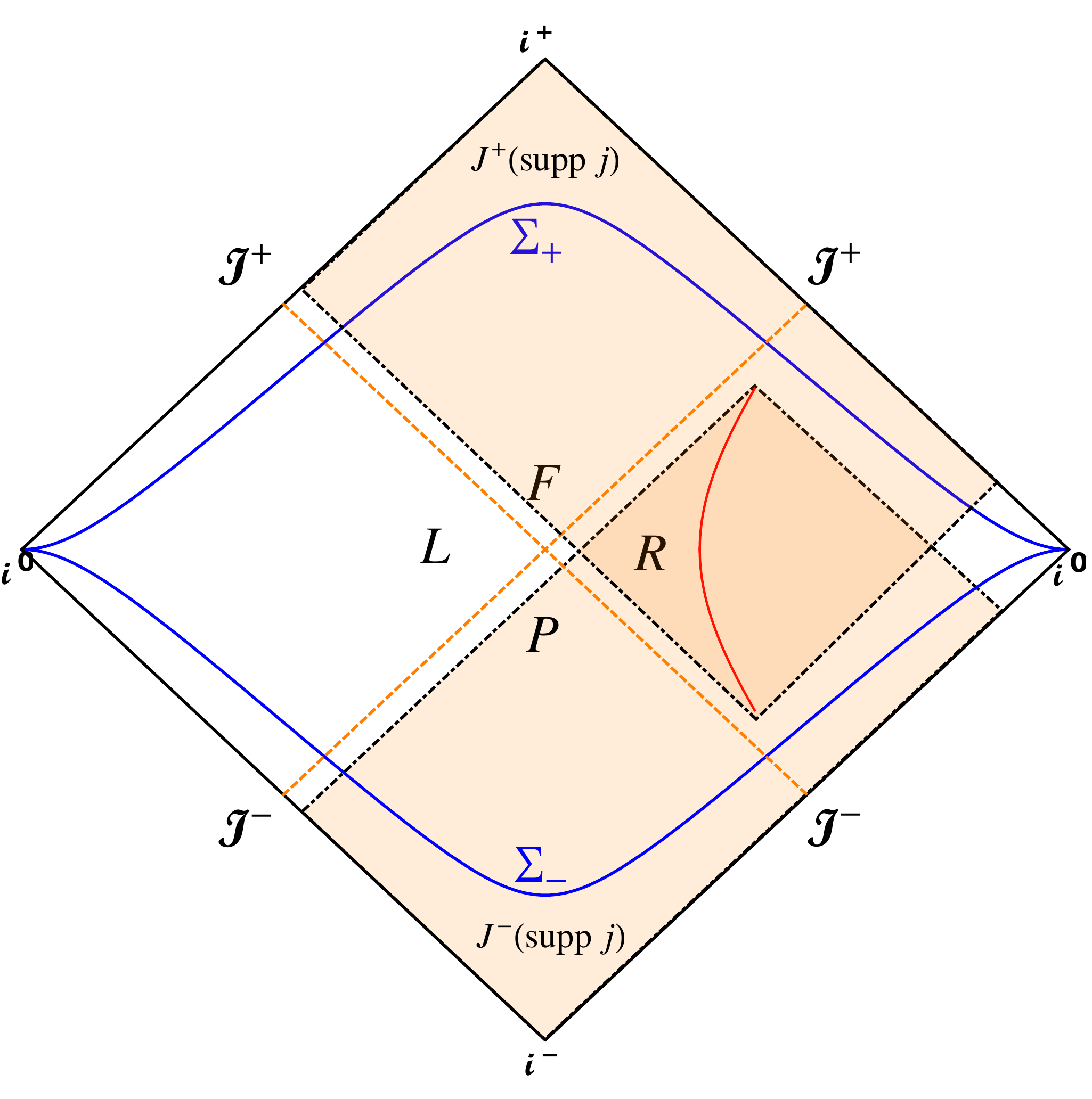}
\end{center}
\caption{The figure shows the conformal Minkowski spacetime with a uniformly accelerated finite-time 
source satisfying $z^2-t^2=1/2$ at the right Rindler wedge~$I$. $\Sigma_\mp$ represent $t=\mp 2$ 
past and future Cauchy surfaces of the Minkowski spacetime. }
\label{fig1}
\end{figure}

The source impact on the classical field $\phi$ is ruled by the inhomogeneous Klein-Gordon equation: 
\begin{equation}
\nabla^a\nabla_a\phi =  j,
\label{sourceKGeq}
\end{equation}
where $\nabla_a$ is the torsion-free covariant derivative compatible with $\eta_{ab}.$
 Let us denote the retarded solution of Eq.~(\ref{sourceKGeq}) by
\begin{equation}
Rj(x)\equiv \int_{\mathbb{R}^4}dx'  \sqrt{-g} \; G_{\rm ret}(x,x')j(x'),
\label{retardedsol}
\end{equation}
where $G_{\rm ret}$ is the retarded Green function for a spacetime point source~\cite{IZ}. 
By choosing a Minkowski space Cauchy surface 
$\Sigma_+ \subset \mathbb{R}^4 - J^-\left({\rm supp}\;j \right)$,  
where ${\rm supp}\;j$ is the support of $j$, we can write the retarded solution on $\Sigma_+$ as 
\begin{equation}
Rj=-(Aj-Rj)\equiv - Ej,
\label{Ej}
\end{equation} 
since the advanced solution, $Aj$, vanishes on  $\mathbb{R}^4 - J^-\left({\rm supp}\;j\right)$ (see Fig.~\ref{fig1}). 

It is important to understand that (for $T\to\infty$) $R = A$ 
inside the right Rindler wedge~\cite{B,RW}.   In that region the solution is 
invariant under time reversal and under Rindler time translation.  
It is not surprising that the temporal Fourier decomposition of 
such a function involves only the frequency $\omega=0$.

Now, let us analyze the radiation emitted by the scalar source from the perspective of 
inertial observers in the asymptotic future. For this purpose, we will decompose 
Eq.~(\ref{Ej}) in $\Sigma_+$ using the so-called Unruh modes~\cite{U} 
$\left\{w^1_{\omega {\bf k}_\bot}, w^2_{\omega {\bf k}_\bot} \right\}$, 
where $\omega\in \mathbb{R}^+$ and ${\bf k}_\perp \in \mathbb{R}^2$. 
They can be written in terms of left- and right-Rindler modes 
$v^R_{\omega {\bf k}_\bot}$ and $v^L_{\omega {\bf k}_\bot}$, respectively, as
\begin{eqnarray}
w^1_{\omega {\bf k}_\perp} & \equiv &
\frac{v_{\omega {\bf k}_\perp}^R + 
e^{-\pi\omega/a}v_{\omega\,-{\bf k}_\perp}^{L*}}
{\sqrt{1-e^{-2\pi\omega/a}}},
\label{wpositive1}\\
w^2_{\omega {\bf k}_\perp} & \equiv &
\frac{v_{\omega {\bf k}_\perp}^L + e^{-\pi\omega/a}v_{\omega\, 
-{\bf k}_\perp}^{R*}}{\sqrt{1-e^{-2\pi\omega/a}}}.
\label{wpositive2}
\end{eqnarray} 
The modes $v_{\omega {\bf k}_\perp}^R$ vanish in the left Rindler wedge 
and take the following form in the right Rindler wedge: 
 \begin{equation}
v_{\omega {\bf k}_\perp}^R =e^{-i\omega \tau} F_{\omega {\bf k}_{\perp}}(\xi, {\bf x}_\perp),
 \label{RRindmode}
 \end{equation}
with 
\begin{equation} 
F_{\omega {\bf k}_{\perp}}(\xi, {\bf x}_\perp)\equiv \left[\frac{\sinh(\pi\omega/a)}{4\pi^4 a}\right]^{1/2}
K_{i\omega/a}\left(\frac{k_\perp}{a}e^{a\xi}\right)e^{i{\bf k}_\perp \cdot {\bf x}_\perp}.
\label{F}
\end{equation}
The modes $v_{\omega {\bf k}_\perp}^L$ are defined by 
$v_{\omega {\bf k}_\perp}^L (t,x,y,z)=v_{\omega {\bf k}_\perp}^{R*} (-t,x,y,-z)$. 
Hence, they vanish in the right Rindler wedge and take the form~(\ref{RRindmode}) 
in Rindler coordinates covering the left Rindler wedge. 

The Unruh modes~(\ref{wpositive1}) and~(\ref{wpositive2}), together with their Hermitian conjugates,
 form a complete orthonormal set  (with respect to the Klein-Gordon inner product~\cite{CHM08}) 
 of solutions of the {\em homogeneous} Klein-Gordon equation with initial data in certain natural Hilbert spaces.  
 Although they are labeled by Rindler energy $\omega$ and transverse momentum ${\bf k}_\perp,$ 
 they are positive-frequency with respect to the inertial time $t$. This makes them particularly suitable to 
 analyze the relation between radiation seen by inertial observers and the physics of uniformly accelerated 
 observers. Mathematically, they provide a factorization of the complicated Bogolubov transformation 
 relating the Rindler modes to the conventional modes in Minkowski space: The mapping (Rindler $\leftrightarrow$ 
 Unruh) mixes positive and negative frequency but does not change the spatial dependence of the 
 functions, while the mapping (Unruh $\leftrightarrow$ Minkowski) does not mix the sign of frequency 
 (particles with antiparticles) but interchanges the Rindler eigenfunctions~(\ref{F}) with the conventional plane waves.

 Using  Eq.~(\ref{Ej})  and the fact that $\nabla^a\nabla_aRj =0$ outside ${\rm supp}\;j$, 
 we can express  $Rj$ on $\Sigma_+$  in terms of the modes $w^\sigma_{\omega {\bf k}_\perp}$ 
 ($\sigma=1,2$) as 
\begin{equation}
Rj=-\sum_{\sigma=1}^{2}\int_0^\infty d\omega 
\int d^2{\bf k}_\perp \langle w^\sigma_{\omega {\bf k}_\perp}, Ej\rangle_{{}_{\rm KG}} w^\sigma_{\omega {\bf k}_\perp} + {\rm c.c.},
\label{retUnrunmodes}
\end{equation}
 where $\langle . \; , \; .\rangle_{{}_{\rm KG}}$ denotes the Klein-Gordon inner product. 
 By means of  the identity~\cite{wald90}
 \begin{equation}
 \langle w^\sigma_{\omega {\bf k}_\perp}, Ej\rangle_{{}_{\rm KG}} 
 = i \int_{\mathbb{R}^4}d^4 x \sqrt{-g} {w}^{\sigma *}_{\omega {\bf k}_\perp}(x) j(x)
 \label{identity0}
 \end{equation}
 together with Eqs.~(\ref{source}) and~(\ref{wpositive1})-(\ref{F}), 
 we can compute the expansion coefficients of Eq.~(\ref{retUnrunmodes}) as 
 \begin{eqnarray}
 \langle w^1_{\omega {\bf k}_\perp}, Ej\rangle_{{}_{\rm KG}} 
 &=&\frac{ i}{\sqrt{1-e^{-2\pi\omega/a}}} \int_{\mathbb{R}^4}d^4 x \sqrt{-g} {v}^{R {}^*}_{\omega {\bf k}_\perp}(x) j(x) 
 \nonumber \\
 &=&\frac{ 2iq}{\sqrt{1-e^{-2\pi\omega/a}}} \frac{\sin(\omega T)}{\omega} F^*_{\omega {\bf k}_{\perp}}(0)
 \label{ampR1}
 \end{eqnarray}
 and 
 \begin{eqnarray}
\langle w^2_{\omega {\bf k}_\perp}, Ej\rangle_{{}_{\rm KG}}\!\!\! 
&=&\!\!\frac{ i e^{-\pi \omega/a}}{\sqrt{1-e^{-2\pi\omega/a}}} 
\int_{\mathbb{R}^4}d^4 x \sqrt{-g} {v}^{R {}}_{\omega -{\bf k}_\perp}(x) j(x) 
\nonumber  \\
&=& \frac{ 2iqe^{-\pi \omega/a}}{\sqrt{1-e^{-2\pi\omega/a}}}   
\frac{\sin(\omega T)}{\omega} F_{\omega -{\bf k}_{\perp}}(0).
\label{ampR2} 
\end{eqnarray}
Using Eq.~(\ref{F}) in Eqs.~(\ref{ampR1}) and~(\ref{ampR2}) 
and taking the limit $T\rightarrow \infty$, where the source is uniformly accelerated forever,
we obtain
 \begin{eqnarray}
 \langle w^1_{\omega {\bf k}_\perp}, Ej\rangle_{{}_{\rm KG}}
 = \langle w^2_{\omega {\bf k}_\perp}, Ej\rangle_{{}_{\rm KG}}
 =\frac{iq K_0(k_\bot/a)}{\sqrt{2\pi^2a}}\delta(\omega),
 \label{amplitudes1}
 \end{eqnarray}
 where we have used that 
 $\lim_{A\rightarrow \infty} \sin(Ax)/x=\pi \delta(x).$  
 Now, using Eq.~(\ref{amplitudes1}) in Eq.~(\ref{retUnrunmodes}) together with the fact that 
 $w^1_{-\omega {\bf k}_\bot}=w^2_{\omega {\bf k}_\bot}$~\cite{CHM08} one obtains
 \begin{equation}
 Rj=-\frac{iq}{\sqrt{2\pi^2a}}\int d^2{\bf k}_\bot K_0(k_\bot/a)w^2_{0{\bf k}_\bot} + c.c. 
 \label{zerofreqexp}
 \end{equation}
The above expansion explicitly shows that, in the limit where the charge accelerates 
forever, only (finite)~\cite{footnote2} {\em zero-energy Unruh modes} contribute to 
the classical radiation seen by inertial observers in the asymptotic future. We also note 
that the expansion amplitudes are built entirely from {\em zero-energy Rindler modes} 
in the right wedge [see Eqs.~(\ref{ampR1})-(\ref{amplitudes1})].  
 
Now, in order to compare our results with the usual results obtained from classical 
(scalar) electrodynamics, let us explicitly compute the integrals in Eq.~(\ref{zerofreqexp}) 
when $Rj$ is evaluated in the region $t>|z|$ of Minkowski spacetime.  [This ``future Rindler wedge''
is also associated with the name of Kasner (in dimension 4) or Milne (in dimension 2).]
 As shown in Ref.~\cite{CHM08}, the Unruh mode $w^2_{\omega {\bf k}_\bot}$ can be written as 
\begin{equation}
w^2_{\omega {\bf k}_{\perp}} 
= -i\frac{e^{i{\bf k}_\perp \cdot {\bf x}_\perp+ i\omega\zeta}}{\sqrt{32\pi^2 a}
}e^{\pi\omega/2a}H^{(2)}_{i\omega/a}\left(k_\perp e^{a\eta}/a\right),
\label{w2F}
 \end{equation}
where $H^{(2)}_\nu$ is the Hankel function of order $\nu$ and we have introduced 
the coordinates $\eta,\zeta \in \mathbb{R}$ defined by 
\begin{equation}
t=a^{-1}e^{a\eta}\cosh a\zeta, \; z=a^{-1}e^{a\eta}\sinh a\zeta,
\label{kasnercoord}
\end{equation} 
which cover the future wedge. By using Eq.~(\ref{w2F}) in Eq.~(\ref{zerofreqexp}) 
we can write the retarded solution $Rj$ as
\begin{equation}
 Rj=\!\frac{-q}{8\pi^2 a}
 \int \!\!d^2{\bf k}_\bot e^{i{\bf k}_\perp \cdot {\bf x}_\perp}K_0(k_\bot/a)H^{(2)}_0\left(\frac{k_\bot}{a} e^{a\eta}\right) 
 + {\rm c.c.}  
 \label{zerofreqexp2}
 \end{equation}
In order to compute this integral, let us define polar coordinates $(k_\perp,\varphi)$ by 
$$k_x=k_\perp \cos\varphi, \: k_y=k_\perp \sin\varphi.$$ In such coordinates, we can 
write ${\bf k}_\perp \cdot {\bf x}_\perp=k_\perp x_\perp \cos\varphi $, 
with $x_\perp\equiv\sqrt{x^2+y^2}$, $d^2 {\bf k}_\perp= k_\perp d\varphi dk_\perp$, and
 \begin{eqnarray}
 Rj 
 &=&\frac{-q}{8\pi^2 a}\int_0^\infty k_\perp dk_\perp\int_0^{2\pi} d\varphi 
 \left[ e^{ik_\perp x_\perp \cos\varphi }K_0(k_\bot/a)\right.  \nonumber \\
 && \left. \times H^{(2)}_0\left(\frac{k_\bot}{a} e^{a\eta}\right)\right] + {\rm c.c.} 
 \label{zerofreqexp3}
 \end{eqnarray}
We can now perform the integral in $\varphi$  by means of the identity 
\begin{equation} 
J_0(\alpha)=\frac{1}{2\pi}\int_0^{2\pi} d\varphi e^{i \alpha\cos\varphi }. 
\end{equation} 
Then, by using $H^{(2)}_0 \equiv J_0 - i Y_0$ we cast Eq.~(\ref{zerofreqexp3}) as 
\begin{equation}
 Rj=\!\frac{-q}{2\pi a}\int_0^\infty dk_\perp k_\perp J_0(k_\perp x_\perp)K_0(k_\bot/a)J_0\left(k_\bot e^{a\eta}/a\right). 
 \label{zerofreqexp4}
 \end{equation}
 Now, by  the equality (see Ref.~\cite{GR} or Eq.~(132) of Ref.~\cite{H17})
 \begin{equation}
 \int_0^\infty dk_\perp k_\perp J_0(k_\perp x_\perp)K_0(k_\bot/a)J_0\left(k_\bot e^{a\eta}/a\right)=a/2\rho_0(x),
 \label{equality}
 \end{equation}
 where $\rho_0(x)\equiv \frac{a}{2}\sqrt{\left(-x^\mu x_\mu + a^{-2}\right)^2 + 4(t^2-z^2)/a^2},$ we find
 the retarded solution $Rj$ to be
 \begin{equation}
 Rj=\frac{-q}{4\pi\rho_0(x)}.   
 \label{RjRW}
 \end{equation}
This is exactly the retarded solution obtained by the usual Green function method of (scalar) 
electrodynamics (see, for instance, Eq.~(2.2) of Ref.~\cite{RW} with $\rho \to -j$).
 
 
 We stress that the foregoing analysis involves no quantum theory whatsoever.  
 The Rindler and Unruh modes enter only as classical special functions describing 
 the structure of the space of classical solutions of the field equation.  
 Furthermore, no ``particle" concept was mentioned and no perturbation theory was employed.  
 Nevertheless, it is possible~\cite{HM} to introduce a notion of ``classical particle number" 
 that provides an illuminating comparison with the quantum calculations to follow:  
 Define the classical number of particles radiated (as seen by inertial observers) to be
 \begin{equation}
 N_M\equiv \langle K Rj, K Rj\rangle_{_{\rm KG}},
 \label{Nclassdef}
 \end{equation}
where
  \begin{equation}
K Rj\equiv -\frac{iq}{\sqrt{2\pi^2a}}\int d^2{\bf k}_\bot K_0(k_\bot/a)w^2_{0{\bf k}_\bot}
 \label{retzero} 
 \end{equation}
 is the (inertial-) 
 positive-frequency part of the retarded 
solution in Eq.~(\ref{zerofreqexp}). Using the orthonormality of the Unruh modes 
 $$ 
 \langle w^{\sigma}_{\omega {\bf k}_\perp},  w^{\sigma'}_{\omega' {\bf k'}_\perp}\rangle_{{}_{\rm KG}}
 =\delta_{\sigma \sigma'}\delta(\omega-\omega')\delta({\bf k}_\bot -{\bf k'}_\bot),
 $$ 
 we substitute Eq.~(\ref{retzero}) in Eq.~(\ref{Nclassdef}) to  obtain
  \begin{eqnarray}
  \frac{N_M}{T_{\rm tot}}&=&\frac{q^2}{2\pi^2 a}\int_{0}^\infty d k_\perp k_\perp|K_0(k_\perp/a)|^2 \nonumber \\
  &=&\frac{q^2 a}{4\pi^2}, 
 \label{NMrate3}
 \end{eqnarray}
where we have used $T_{\rm tot} = 2 \pi \delta(\omega)|_{\omega=0}$ and $$\int_0^\infty dx  \; x |K_0(x)|^2=1/2.$$ 
This is exactly the result obtained in \cite{RW} using tree-level QFT. We will reobtain this result nonperturbatively in the context of 
QFT in the next section.

\section{Radiation emitted by an accelerated charge: Quantum aspects}
\label{sec:III}

Let us now analyze the radiation emitted by the charge from the 
point of view of QFT. To unveil the radiation 
content seen by an inertial observer in the asymptotic future, we 
now analyze the impact of the source~(\ref{source}) (which is 
still a c-number and a scalar) on a 
{\em 
quantum} scalar field $\hat{\phi}$ satisfying
\begin{equation}
\nabla^a\nabla_a\hat{\phi} = j. 
\label{quantumsourceKGeq}
\end{equation}
We can write a general solution of this operator equation as 
 \begin{equation}
 \hat{\phi}(t,{\bf x})= Rj(t, {\bf x}) \hat{I} + \hat{\phi}_{\rm in}(t, {\bf x}),
 \label{infield}
 \end{equation}
 where $Rj$ is given by Eq.~(\ref{retardedsol}) and $\hat{\phi}_{\rm in}$ 
satisfies the free (homogeneous) Klein-Gordon equation
 \begin{equation}
 \nabla^a\nabla_a\hat{\phi}_{\rm in}=0.
 \end{equation}
 As a result, we can expand $\hat{\phi}_{\rm in}$ as 
 \begin{equation}
\hat{\phi}_{\rm in}(t,{\bf x})
\equiv \sum_j \left[u_j(t,{\bf x}) \hat{a}_{\rm in}({u}^*_j)
+{u}^*_j(t,{\bf x}) \hat{a}_{\rm in}^\dagger(u_j)\right], 
\label{infieldexp}
\end{equation}
 where $\{u_j\}$ is a set of (Minkowski) positive-frequency modes 
(which might be either  plane waves or Unruh modes). 
Then  $|0^M_{\rm in}\rangle$ (satisfying  $\hat{a}_{\rm 
in}(u^*_j)|0^M_{\rm in}\rangle=0$ for all $j$) is the vacuum 
state as defined by inertial observers in the asymptotic past, and 
the Fock space built from the action of $\hat{a}_{\rm 
in}^\dagger(u_j)$ on $|0^M_{\rm in}\rangle$ describes particle 
states seen by such observers.
 
Alternatively, we can write a solution of Eq.~(\ref{quantumsourceKGeq}) as 
 \begin{equation}
 \hat{\phi}(t,{\bf x})=Aj(t, {\bf x}) \hat{I} + \hat{\phi}_{\rm out}(t, {\bf x}),
 \label{outfield}
 \end{equation}
 where  we recall that $Aj$ is the advanced solution of Eq.~(\ref{sourceKGeq})  
 [which vanishes on $ \mathbb{R}^4 - J^- \left({\rm supp}\; j\right)$] and 
$\hat{\phi}_{\rm out}$ satisfies the free Klein-Gordon equation 
 \begin{equation}
 \nabla^a\nabla_a\hat{\phi}_{\rm out}=0.
 \end{equation}
 Therefore, the field $\hat{\phi}_{\rm out}$ can be expanded as 
 \begin{equation}
\hat{\phi}_{\rm out}(t,{\bf x})
\equiv \sum_j \left[v_j(t,{\bf x}) \hat{a}_{\rm out}({v}^*_j)+{v}^*_j(t,{\bf x}) \hat{a}_{\rm out}^\dagger(v_j)\right], 
\label{outfieldexp}
\end{equation}
where $\{v_j\}$ is again 
a set of (Minkowski) positive-frequency modes 
(possibly the same as the $u_j$).   
Hence, $|0^M_{\rm out}\rangle$ (satisfying  $\hat{a}_{\rm out}(v^*_j)|0^M_{\rm out}\rangle=0$ for all $j$)  
is the vacuum state as defined by inertial observers in the asymptotic future and the Fock space built from  
the action of $\hat{a}_{\rm out}^\dagger(v_j)$ on $|0^M_{\rm out}\rangle$ describes particle states seen 
by such observers. 

We can exactly connect the in and out Fock spaces by means of the S-matrix~(see, e.g., Sec. 4.1.4 of Ref.~\cite{IZ}): 
\begin{equation}
\hat{S}= \exp{\left[-i\int d^4x\sqrt{-g} \; \hat{\phi}_{\rm out}(x) j(x)\right]}. 
\label{S}
\end{equation} 
 Let us suppose now that the field is prepared in the in-vacuum, 
$|0^M_{\rm in}\rangle$.  
Then $\hat{\phi}_{\rm int}$ can be identified with 
$\hat{\phi}_{\rm out}$, and  
the S-matrix~(\ref{S})  relates $|0^M_{\rm in}\rangle$ and
$|0^M_{\rm out}\rangle$ as 
\begin{equation}
|0^M_{\rm in}\rangle=\hat{S}|0^M_{\rm out}\rangle.
\label{in-out}
\end{equation}
 Next, we use Eq.~(\ref{in-out}) to analyze how the in-vacuum is seen by inertial observers 
 in the asymptotic future and, as a result, study the radiation emitted by the source.  
 In order to do this, let us first expand $\hat{\phi}_{\rm out}$ in terms of Unruh 
 modes~(\ref{wpositive1}) and~(\ref{wpositive2}):
 \begin{eqnarray}
 \hat{\phi}_{\rm out}(x)
 &=& \sum_{\sigma=1}^{2} \int_0^\infty d\omega \int d^2{\bf k}_\perp 
 \left[w^\sigma_{\omega {\bf k}_\perp}(x) \hat{a}_{\rm out}(w^{\sigma *}_{\omega {\bf k}_\perp})\right. 
 \nonumber \\
 &+&  \left. w^{\sigma *}_{\omega {\bf k}_\perp}(x) \hat{a}_{\rm out}^\dagger(w^\sigma_{\omega {\bf k}_\perp})\right].
 \label{unruhout}
 \end{eqnarray}
 Now, we smear Eq.~(\ref{unruhout}) with $j$, getting
  \begin{eqnarray}
-i\hat{\phi}_{\rm out}(j)\!\!\!
&\equiv&
-i \int_{{\mathbb R}^4} d^4x \sqrt{-g} \; \hat \phi_{\rm out}(x) j(x)
\nonumber \\
&=&\!\!\!\sum_{\sigma=1}^{2} \int_0^\infty \!\!\!\! d\omega \int \!d^2{\bf k}_\perp 
\left[ \langle w^\sigma_{\omega {\bf k}_\perp}, Ej\rangle^*_{{}_{\rm KG}}  \hat{a}_{\rm out}(w^{\sigma *}_{\omega {\bf k}_\perp}) \right. 
\nonumber \\
 &-& \left. \langle w^\sigma_{\omega {\bf k}_\perp}, Ej\rangle_{{}_{\rm KG}}  \hat{a}_{\rm out}^\dagger(w^\sigma_{\omega {\bf k}_\perp})\right],
 \label{unruhoutsmeared}
 \end{eqnarray}
where we have used Eq.~(\ref{identity0}) to transform the spacetime integral 
into the Klein-Gordon inner product.  The (inertial-) 
positive-frequency part of $Ej$ is given by
\begin{equation}
KEj\equiv  
\sum_{\sigma=1}^{2} \int_0^\infty d\omega \int d^2{\bf k}_\perp  
\langle w^\sigma_{\omega {\bf k}_\perp}, Ej\rangle_{{}_{\rm KG}}  w^\sigma_{\omega {\bf k}_\perp},
\label{KEj}
\end{equation}
with which we can rewrite Eq.~(\ref{unruhoutsmeared}) as 
 \begin{equation}
 i\hat{\phi}_{\rm out}(j)=  \hat{a}^\dagger_{\rm out}\left(KEj\right) -\hat{a}_{\rm out}\left({KEj}^* \right),
 \label{unruhoutsmeared2}
\end{equation}
where we have used 
\begin{equation}
\hat{a}^\dagger_{\rm out}\left(KEj\right)
\!=\!\sum_{\sigma=1}^2 \int_0^\infty \!\!\!d\omega \!\int \!\!d{\bf k}_\perp  
\langle w^{\sigma}_{\omega {\bf k}_\perp}, Ej\rangle_{{}_{\rm KG}}\hat{a}^\dagger_{\rm out}(w^{\sigma}_{\omega {\bf k}_\perp})
\label{adagger}
\end{equation}
and analogously for $\hat{a}_{\rm out}\left({KEj}^* \right)$. This enables us to cast the S-matrix as 
\begin{equation}
\hat{S}=\exp{\left[\hat{a}_{\rm out}\left(KEj^* \right)- \hat{a}^\dagger_{\rm out}\left(KEj\right)\right]}.
\label{S2}
\end{equation}
(In Eqs.~(\ref{unruhoutsmeared2}) and (\ref{S2}), ${KEj}^*$ is to be 
read as $(KEj)^*$; we prefer not to overload the notation with an 
extra pair of parentheses.)
Applying the  Zassenhaus formula, 
$$e^{\mathfrak{a}+\mathfrak{b}}=e^{\mathfrak{a}}e^{\mathfrak{b}}e^{-\frac{1}{2}[\mathfrak{a},\mathfrak{b}]}$$  
when 
$[\mathfrak{a},\mathfrak{b}]$ is a c-number, we get 
from Eqs.\ (\ref{in-out}) and (\ref{S2}) 
the following expression for the in-vacuum in terms of 
out-states:
\begin{equation}
|0^M_{\rm in}\rangle=e^{-\| KEj\|^2/2}e^{-\hat{a}^\dagger_{\rm out}\left(KEj\right)}|0^M_{\rm out}\rangle,
\label{in-out2}
\end{equation}
 where 
\begin{eqnarray}
 \left[\hat{a}_{\rm out} \left(KEj^*\right), 
\hat{a}^\dagger_{\rm out}\left(KEj\right) \right]
&= &\langle KEj, KEj\rangle_{{}_{\rm KG}} \hat{I} 
\nonumber \\
&\equiv&  \| KEj\|^2 \hat{I}.  
\nonumber
\end{eqnarray}
This is a (multi-mode) coherent state, i.e., it is an eigenstate of 
$\hat{a}_{\rm out}(w^{\sigma *}_{\omega {\bf k}_\perp})$ 
with eigenvalue 
$ -\langle w^{\sigma}_{\omega {\bf k}_\perp}, Ej\rangle_{{}_{\rm KG}}$ 
for all $\sigma\in \{1,2\}$, $\omega\in \mathbb{R}^+$, and ${\bf k}_\perp \in \mathbb{R}^2.$ 
To see this, we note that
\begin{widetext}
  \begin{eqnarray}
 \hat{a}_{\rm out}(w^{\sigma *}_{\omega {\bf k}_\perp})|0^M_{\rm in}\rangle
 &=& e^{-\| KEj\|^2/2}e^{-\hat{a}^\dagger_{\rm out}
 \left(KEj\right)}\left( e^{\hat{a}^\dagger_{\rm out}\left(KEj\right)}\hat{a}_{\rm out}(w^{\sigma *}_{\omega {\bf k}_\perp})
 e^{-\hat{a}^\dagger_{\rm out}\left(KEj\right)}\right)|0^M_{\rm out}\rangle 
 \nonumber \\
 &=& e^{-\| KEj\|^2/2}e^{-\hat{a}^\dagger_{\rm out}\left(KEj\right)} \left(\hat{a}_{\rm out}(w^{\sigma *}_{\omega {\bf k}_\perp})
 + \left[\hat{a}^\dagger_{\rm out}\left(KEj\right), \hat{a}_{\rm out}(w^{\sigma *}_{\omega {\bf k}_\perp}) 
 \right]\right)|0^M_{\rm out}\rangle 
 \nonumber \\
 \end{eqnarray}
 \end{widetext}
 where we have used in the second line that 
  \begin{equation}
 e^{\hat{A}}\hat{B}e^{-\hat{A}}=\hat{B} +[\hat{A}, \hat{B}] 
+\frac{1}{2!}[\hat{A},[\hat{A},\hat{B}]] + \cdots. 
 \label{identity}
 \end{equation} 
   Using 
 $$
 \left[\hat{a}_{\rm out}(w^{\sigma *}_{\omega {\bf k}_\perp}), \hat{a}^\dagger_{\rm out}\left(KEj\right) \right]
 = \langle w^{\sigma}_{\omega {\bf k}_\perp}, KEj\rangle_{{}_{\rm KG}} \hat{I}
 $$
 and $\hat{a}_{\rm out}(w^{\sigma *}_{\omega {\bf k}_\perp})|0^M_{\rm out}\rangle=0,$ we find  that
 \begin{eqnarray}
 \hat{a}_{\rm out}(w^{\sigma *}_{\omega {\bf k}_\perp})|0^M_{\rm in}\rangle
 = -\langle w^{\sigma}_{\omega {\bf k}_\perp}, Ej\rangle_{{}_{\rm KG}}|0^M_{\rm in}\rangle,
 \label{coherent1}
 \end{eqnarray}
 where we have also 
used Eq.~(\ref{in-out2}) and  $\langle 
w^{\sigma}_{\omega {\bf k}_\perp}, KEj\rangle_{{}_{\rm KG}}=\langle w^{\sigma}_{\omega {\bf k}_\perp}, Ej\rangle_{{}_{\rm KG}}$.  
 
From  Eqs.~(\ref{unruhout}) and~(\ref{coherent1}), it is 
easy to see that 
 \begin{eqnarray}
\langle 0^M_{\rm in}| \hat{\phi}_{\rm out}|0^M_{\rm in}\rangle\!\!\!
&=&\!\!\!-\sum_{\sigma=1}^2\int_0^\infty \!\!\!\!\! d\omega \int \!\!d^2{\bf k}_\perp 
\langle w^{\sigma}_{\omega {\bf k}_\perp}, Ej\rangle_{{}_{\rm KG}} w^{\sigma}_{\omega {\bf k}_\perp} 
\nonumber \\ 
&+& {\rm c.c.},
 \label{coherent2}
 \end{eqnarray}
 which is just $-Ej$ written in terms of Unruh modes. 
By evaluating the above expression on 
 $\Sigma_+$, where  $Aj=0$,  
we have that $-Ej=Rj$ and thus (in the far future) 
  \begin{eqnarray}
\langle 0^M_{\rm in}| \hat{\phi}_{\rm out}(x) |0^M_{\rm in}\rangle = Rj(x),
 \label{coherent3}
 \end{eqnarray}
which is the classical retarded solution produced by the uniformly accelerated source. 
Therefore, for such a source interacting with a quantum field $\hat{\phi}$ in the vacuum 
state defined by inertial observers in the asymptotic past, inertial observers in the asymptotic 
future will describe this state as a  coherent state  with field expectation value given by $Rj,$ 
the classical retarded solution~\cite{footnote2}.  

As we have stressed previously, this solution is built 
exclusively from zero-energy Rindler modes.  
Our next task is to show that this property extends  not only 
to the field expectation value but to  the quantum state as a whole. 
 Let us  delve into the structure of the 
coherent state~(\ref{in-out2}) to see how each Unruh mode contributes to it 
in the limit $T_{\rm  tot}\rightarrow \infty$. 
Use Eqs.~(\ref{ampR1})--(\ref{zerofreqexp}) 
to recast 
Eq.~(\ref{adagger}) as 
\begin{equation}
\!\hat{a}^\dagger_{\rm out}\left(KEj\right)
=\frac{iq}{\sqrt{2\pi^2a}}\int d^2{\bf k}_\bot K_0(k_\bot/a)\hat{a}^\dagger_{\rm out}\left(w^2_{0{\bf k}_\bot}\right). 
\label{quantumKEj}
\end{equation}
Using Eq.~(\ref{quantumKEj}) in Eq.~(\ref{in-out2}) one obtains 
\begin{eqnarray}
& & | 0_{\rm in}^M\rangle = \exp \left[-T_{\rm tot}q^2a/(4\pi^2)\right]
\nonumber \\
& &\bigotimes_{{\bf k}_\perp}
 \exp{\!\!\left[\frac{iq K_0({k_\perp}/{a})}{\sqrt{2\pi^2a}}\hat{a}_{\rm out}^\dagger\left(w^2_{0{\bf k}_\bot}\right)\right]}\!|0_{\rm out}^M\rangle,
\label{coherentzerofreq}
\end{eqnarray}
where we have used that 
\begin{eqnarray}
\|KEj\|^2
&=& \sum_{\sigma=1}^2 \int_0^\infty d\omega \int d^2{\bf k}_\perp |\langle w^\sigma_{\omega {\bf k}_\perp}Ej \rangle_{_{\rm KG}}|^2
\nonumber \\
&=&q^2a T_{\rm tot}/4 \pi^2,
\label{KEfsquared}
\end{eqnarray} 
which comes from using Eq.~(\ref{amplitudes1}) in the above equation 
together with $T_{\rm tot} = 2\pi \delta (\omega)|_{\omega=0}$
(as usual in quantum scattering theory).   
Thus, we see from Eq.~(\ref{coherentzerofreq}) that only zero-energy 
Unruh modes contribute to build the quantum radiation emitted by the charge 
when the field is initially in the Minkowski vacuum state. 
This result vindicates the claim that {\em each particle emitted
in the inertial frame must correspond in the accelerated
one to either the emission or the absorption of a zero-energy 
Rindler particle}~\cite{HMS92},\cite{RW}.


Let us finish the analysis of the quantum aspects of the 
radiation by studying both
the mean number of created particles and  
 the expectation value of the stress-energy tensor in the 
asymptotic future. This will allow us to 
 clarify further the 
connection between the quantum and classical radiation emissions. 
We begin by  using the fact 
that $|0_{\rm in}^M\rangle$ is 
a coherent state for inertial observers in the asymptotic 
future --- Eq.~(\ref{coherent1}) --- 
to calculate the mean particle number in each Unruh mode:
 \begin{eqnarray}
 \langle 0_{\rm in}^M| \hat{N}^{\rm out}_{\omega {\bf k}_\perp} | 0_{\rm in}^M\rangle 
 &=&| \langle w^{\sigma}_{\omega {\bf k}_\perp}, Ej\rangle_{{}_{\rm KG}}| ^2, 
 \label{quantumN1}
 \end{eqnarray}
 where $ \hat{N}^{\rm out}_{\omega {\bf k}_\perp} = \hat{a}^\dagger_{\rm out}
(w^{\sigma}_{\omega {\bf k}_\perp}) \hat{a}_{\rm out}(w^{\sigma *}_{\omega 
{\bf k}_\perp})$ and  we have used  
$\langle 0_{\rm in}^M| 0_{\rm in}^M\rangle=1.$ 
Integrating over all quantum numbers, we find that the total mean 
particle number created at the asymptotic future is
\begin{eqnarray}
 \langle 0_{\rm in}^M| \hat{N}^{\rm out} | 0_{\rm in}^M\rangle
 &=& \sum_{\sigma=1}^2 \int_0^\infty d\omega 
 \int d^2{\bf k}_\perp |\langle w^\sigma_{\omega {\bf k}_\perp}Ej \rangle_{_{\rm KG}}|^2. 
 \nonumber \\
 \label{quantumN2}
 \end{eqnarray} 
Using Eq.~(\ref{KEfsquared}) in this equation one obtains 
\begin{equation}
\frac{\langle 0_{\rm in}^M| \hat{N}^{\rm out}| 0_{\rm in}^M\rangle}{T_{\rm tot}}= \frac{q^2 a}{4\pi^2},
 \label{NMrate4}
 \end{equation}
 which agrees with the classical particle number per proper time given in Eq.~(\ref{NMrate3}). 
 
 Moreover, the agreement between the classical and quantum observables happens not 
 only for the number of emitted scalar particles but also for the (normal-ordered) stress-energy tensor: 
  \begin{eqnarray}
: \hat{T}^{\rm out}_{ab}: \; \equiv \hat{T}^{\rm out}_{ab}-\langle 0_{\rm out}^M|  \hat{T}^{\rm out}_{ab} | 0_{\rm out}^M\rangle,
 \label{normalTab}
 \end{eqnarray}
 where 
 $
 \hat{T}^{\rm out}_{ab}
 =\nabla_a\hat{\phi}_{\rm out}\nabla_b\hat{\phi}_{\rm out}-\frac{1}{2}\eta_{ab}\nabla^c\hat{\phi}_{\rm out}\nabla_c\hat{\phi}_{\rm out}
 $.
To show this, let us use Eqs.~({\ref{unruhout}),~(\ref{coherent1}), 
and~(\ref{retUnrunmodes}) to straightforwardly compute (in the asymptotic future) 
\begin{eqnarray}
\langle 0_{\rm in}^M|:\nabla_a\hat{\phi}_{\rm out}\nabla_b\hat{\phi}_{\rm out} : | 0_{\rm in}^M\rangle = \nabla_a Rj \nabla_b Rj.
 \label{inter1}
 \end{eqnarray}
  It follows by    
 Eqs.~(\ref{normalTab})--(\ref{inter1}) that  on 
$\mathbb{R}^4 - J^-\left({\rm supp}\;j\right)$ 
we have  
\begin{eqnarray}
\langle 0_{\rm in}^M|: \hat{T}^{\rm out}_{ab}:| 0_{\rm 
in}^M\rangle \equiv \nabla_aRj  \nabla_bRj -
\frac{1}{2}\eta_{ab}\nabla^c Rj \nabla_c Rj,  
 \nonumber \\ 
 \label{<normalTab>}
 \end{eqnarray}
which is precisely the {\rm classical} stress-energy tensor associated with the 
retarded solution, $T_{ab}[Rj]$. As a result, any observable calculated from 
the expectation value of the {\rm quantum} stress-energy tensor in the (Minkowski) 
in-vacuum will be given by its classical counterpart computed from $T_{ab}\left[Rj\right]$. 
In particular, the (quantum) energy flux integrated along a large sphere in the asymptotic 
future will be given by~\cite{RW}
\begin{equation}
\int d S^b  \langle 0_{\rm in}^M|: \hat{T}^{\rm out}_{ab}:| 0_{\rm in}^M\rangle (\partial_t)^a = \frac{q^2a^2}{12\pi},
\label{scalarlarmor}
\end{equation}
which is the usual Larmor formula for the power radiated by a scalar source 
(with respect to inertial observers). Here, $d S^b$ is the vector-valued volume 
element on the sphere and $(\partial_t)^a$ is the Killing field associated with 
a global inertial congruence.

}

\section{Conclusions} \label{sec:IV} 
We have analyzed both 
classical and quantum aspects of the 
radiation emitted by a uniformly accelerated scalar source,
with emphasis on the limit where the lifetime $T_{\rm tot}$ of 
the source is infinite.  
In that case we have shown that the classical radiation is entirely built from 
zero-energy Unruh modes and that only zero-energy Rindler modes 
in the right Rindler wedge contribute to the expansion 
amplitudes. By studying the quantum evolution of the scalar field 
interacting with a classical uniformly accelerated source, we 
were able to show how the quantum and classical analyses relate 
to each other.  When the field is prepared in the vacuum state 
for inertial observers in the asymptotic past, inertial observers 
in the asymptotic future will describe this state as a coherent 
multiparticle state whose field expectation value is given by the classical 
retarded solution $Rj$. This coherent state is built only from 
zero-energy Unruh modes (agreeing with the classical analysis) 
and, as in the classical case, only zero-energy Rindler modes in 
the right Rindler wedge contribute to the amplitudes building the 
coherent state. It is important to stress that this 
 happens only because the field state is the Minkowski in-vacuum. By computing 
the mean particle number and the expectation value of the 
stress-energy tensor in the asymptotic future, we were able to 
show that they agree with their classical counterparts calculated 
with the retarded solution.  As a result, any (quantum) 
observables related to the number of particles or stress-energy 
tensor can be computed using the classical retarded solution. 
More importantly, zero-energy Rindler modes are not a 
mathematical artifact; they play a crucial role in building the 
radiation both in the classical and in the quantum realm. We 
believe that our results put to rest any doubts questioning the 
relationship between the Unruh effect and the classical Larmor 
radiation.

\acknowledgments
A.~L. and S.~F. would like to acknowledge partial support from the Texas A \& M University (TAMU) 
and Sao Paulo Research Foundation (FAPESP) agreement under grant 2017/50388-6 and from the Institute 
for Quantum Science and Engineering at TAMU. A.~L. and G.~M. were also partially supported by FAPESP
under grant 2017/15084-6 and  Conselho Nacional de Desenvolvimento Cient\'\i fico e Tecnol\'ogico, respectively.
The authors are also thankful to Bill Unruh, Don Page, and Bob Wald for helpful discussions. 


\begin{thebibliography}{99}
\bibitem{footnote} In order to avoid the appearance of divergences in dealing 
with finite-time sources, one must implicitly assume that the source is switched 
on/off continously. This is not a source of concern here because we take the limit 
$ T\rightarrow \infty$ before radiation rates are evaluated.
\bibitem{footnote2} We note that  $w^\sigma_{0 k_\bot}$ are well 
defined through the limit $w^\sigma_{0 k_\bot} =\lim_{\omega\to 0} w^\sigma_{\omega k_\bot}$.
\bibitem{footnote2}
We note that Eq.~(\ref{coherent3}) is still valid for any compact support source $j$ provided that the 
in state is in the Minkowski vacuum.
\end{thebibliography}
\end{document}